**Role of impurity oxygen in superconductivity of "non-doped" $T'$-(La,$RE$)$_2$CuO$_4$**

A. Tsukada [a,*], M. Noda [a,b], H. Yamamoto [a], M. Naito [c]

[a] NTT Basic Research Laboratories, NTT Corporation, 3-1 Morinosato-Wakamiya, Atsugi, Kanagawa 243-0198, Japan

[b] Tokyo University of Science, 2641 Yamazaki, Noda, Chiba 278-8510, Japan

[c] Department of Applied Physics, Tokyo University of Agriculture and Technology, 2-24-16 Naka-cho, Koganei, Tokyo 184-8588, Japan

**Abstract**

We have systematically investigated the effect of oxygen nonstoichiometry in a nominally undoped superconductor $T'$-(La,Y)$_2$CuO$_{4+y}$. In the experiments, the reduction condition was changed after the sample growth by MBE. The superconductivity is very sensitive to the reduction condition. With systematically increasingly reduced atmospheres, resistivity shows a continuous drop and no discontinuity is observed even until the appearance of superconductivity. The absence of the highly insulating state expected for Mott insulators around $y \sim 0$ suggests that $T'$-(La,Y)$_2$CuO$_4$ has intrinsic carriers. The role of residual apical oxygen, which is detrimental to superconductivity, is also discussed based on the resistivity-temperature characteristics in insufficiently reduced samples.

*Corresponding author.

Dr. Akio Tsukada

Postal address: Thin-Film Materials Research Group, NTT Basic Research Labs., 3-1 Wakamiya Morinosato, Atsugi-shi, Kanagawa 243-0198, Japan

Phone: +81-46-240-3349

## 1. Introduction

Superconductivity in cuprates with the $Nd_2CuO_4$ (so-called $T'$) structure is very sensitive to impurity oxygen at O(3) sites (apical sites), whose presence is very detrimental to superconductivity. It is well known that as-grown $Pr_{1.85}Ce_{0.15}CuO_4$ or $Nd_{1.85}Ce_{0.15}CuO_4$ specimens, which contain a fair amount of impurity oxygen, are never superconducting [1, 2]. In addition, antiferromagnetic (AF) correlations exist in as-grown non-superconducting specimens, but they essentially disappear in reduced superconducting samples [3, 4]. This implies that more complete removal of impurity oxygen weakens AF correlations and thereby expands the superconducting region. It has actually been demonstrated by Brinkman *et al.* that an enhanced oxygen reduction process in $Pr_{2-x}Ce_xCuO_{4+y}$ indeed expands the superconducting region down to a doping level of around $x = 0.04$ even with slightly increasing critical temperature [5]. Furthermore, very recently, we have discovered superconductivity by isovalent $RE^{3+}$ doping in $T'$-$La_2CuO_4$ ($RE$ = rare earth element) [6-8]. This discovery seems to imply that superconductivity can be achieved in the really undoped state of $T'$-cuprates by complete removal of apical oxygen, but there is a possibility that superconductiviy *might* be achieved by effective electron-doping via oxygen deficiencies. In order to examine the latter possibility, we have performed a systematic study of the oxygen nonstoichiometry effect in $T'$-$(La,RE)_2CuO_{4+y}$.

## 2. Experimental

$La_{1.85}Y_{0.15}CuO_{4+y}$ thin films were grown in a customer-designed molecular beam epitaxy chamber from metal sources using multiple electron-gun evaporators. The details of our MBE growth are described elsewhere [9]. Briefly, the film growth was typically performed at the substrate temperature of ~ 650°C with 1 - 5 sccm of ozone gas (10% $O_3$ concentration). The film thickness was typically ~ 900 Å. We used $YAlO_3$ (100) substrates in this study. After the growth, the films were *reduced* in low partial oxygen pressure to remove interstitial apical oxygen. In order to investigate the oxygen nonstoichiometry effect, the *reduction* parameters were systematically varied: *reduction* temperature ($T_{red}$) from 570 to 650°C; *reduction* time ($t_{red}$) from 0 to 60 minutes; *reduction* atmosphere (partial oxygen pressure, $P_{O2}$) from $10^{-4}$ to $10^{-8}$ Torr. In addition, some films were oxidized by ozone.

## 3. Results and discussion

Figure 1 shows the temperature ($T$) dependence of resistivity ($\rho$) for $La_{1.85}Y_{0.15}CuO_{4+y}$ films with different $T_{red}$. For these films, $P_{O2}$ and $t_{red}$ were fixed at $10^{-8}$ Torr and 10 min, respectively. The superconducting transition temperature ($T_c$) is maximized to $T_c$ ~ 23 K with $T_{red}$ = 600°C. The film with lower $T_{red}$ (570°C) has a depressed $T_c$ ~ 16 K and substantially higher resisivity with low-temperature upturn, whereas the films with higher $T_{red}$ (630°C and 650°C) also have a depressed $T_c$ ~ 20 K and 18 K in spite of there being almost no or only a slight increase in resistivity.

Figure 2 shows the temperature dependence of resistivity for $La_{1.85}Y_{0.15}CuO_{4+y}$ films with different $t_{red}$. For these films, $P_{O2}$ and $T_{red}$ were fixed at $10^{-8}$ Torr and 600°C, respectively. There is a broad maximum (~ 23 K) at $t_{red}$ = 10 min. The $T_c$ and also resistivity do not change much with changing $t_{red}$ from 5 to 20 min. The film with shorter $t_{red}$ (1 min) shows no significant change in the value of room-temperature resistivity, but starts to show upturn in resistivity at low temperatures and has a depressed $T_c$, and the film with the shortest $t_{red}$ of ~ 0 min shows significantly higher resistivity with prominent upturn at low temperatures and no superconductivity. It should be noted that both films with short $t_{red}$ of 0 and 1 min have nearly the same slope of the $\rho$-$T$ curves as the films with optimum $t_{red}$ of 5 and 20 min in the temperature region between 200 and 300 K, indicating that the difference in resistivity of the films with the short and optimum $t_{red}$ is most likely caused by magnetic impurity scattering. On the other hand, the film with prolonged $t_{red}$ of 60 min is qualitatively different, and shows no superconductivity and high resistivity, but no upturn at low temperatures.

Figure 3 shows the temperature dependence of resistivity for $La_{1.85}Y_{0.15}CuO_{4+y}$ films with different $P_{O_2}$. The $T_{red}$ and $t_{red}$ were fixed at 630°C and 10 min, respectively. The samples with $P_{O_2} = 10^{-8}$ -$10^{-7}$ Torr are almost identical and show superconductivity with $T_c > 20$ K, whereas the samples with $P_{O_2} \geq 10^{-6}$ Torr are not superconducting and show monotonically increasing resistivity as $P_{O_2}$ increases, but with metallicity ($d\rho/dT > 0$) maintained above 150 K. Finally, the samples oxidized in ozone pressure show semiconducting behavior ($d\rho/dT < 0$). The ozone-oxidized films certainly have $y > 0$, since the same procedure for $T$-$La_2CuO_{4+y}$ in the same MBE chamber leads to superconductivity at 56 K, indicating $y$ of $T$-$La_2CuO_{4+y}$ is larger than 0.1 [10]. The results in Fig. 3 confirm that excess oxygen atoms in the $T'$-structure do not provide hole carriers as effective dopants to the $CuO_2$ layers, which is in contrast to excess oxygen atoms at the tetrahedral site in $T$-$La_2CuO_{4+y}$. Instead, excess oxygen atoms occupy the interstitial apical sites [O(3)], where it acts only as a pair breaker and strong scatterer [11]. Figure 4 shows the corresponding lattice constants ($a_0$ and $c_0$) for $La_{1.85}Y_{0.15}CuO_{4+y}$ films with different $P_{O_2}$. The $c_0$ shows a very slight but steady decrease with decreasing $P_{O_2}$ in *reduction*, confirming that interstitial oxygen atoms are more completely removed with lower $P_{O_2}$.

The results shown in Fig. 1 to 3 confirm that superconductivity in $La_{1.85}Y_{0.15}CuO_{4+y}$ is very sensitive to oxygen nonstoichiometry. The superconductivity degrades with either insufficient *reduction* (low $T_{red}$, short $t_{red}$, or high $P_{O2}$) or excessive *reduction* (high $T_{red}$ or long $t_{red}$), although it degrades in a different way. In either case, $T_c$ is suppressed and the resistivity increases more or less. But the distinguishable feature between insufficient and excessive *reduction* is the behavior of low-temperature resistivity: the upturn in resistivity is present in the films with insufficient *reduction* and absent with excessive *reduction*. The results can be interpreted as follows. In insufficient *reduction*, removal of impurity oxygen at the apical site is not complete. We speculate that the low-temperature upturn in resistivity seen in this case is due to magnetic impurity (Kondo) scattering, which seems to be caused by Cu spins induced just beneath apical oxygen [12, 13]. On the other hand, in excessive reduction, the removal of apical oxygen is almost complete, but oxygen in the $CuO_2$ plane [O(1)] starts to come out. Further reduction eventually leads to irreversible decomposition. It may be argued that removal of oxygen in the $(La,RE)_2O_2$ layers [O(2)], instead O(1), *might* lead to a similar suppression of $T_c$ by overdoping. If this *were* the case, the resistivity *would* also be lowered by overdoping. However, our experimental results indicate that the suppression in $T_c$ accompanies the increase in resistivity. Hence, we think that the presence of oxygen deficiencies at O(1) is the reason for the degradation of superconductivity in excessive reduction. It seems that the difference in the binding energies for O(1) and O(3) is subtle, which makes the complete removal of O(3) with O(1) intact difficult.

Our present investigations of the oxygen nonstoichiometry effect in $T'$-$(La,RE)_2CuO_{4+y}$ indicate that electron doping via oxygen deficiencies at O(2) is not a source of superconductiviy in nominally undoped $T'$-(La,$RE$)$_2$CuO$_4$. This is because with a fixed composition, the highest $T_c$ can be achieved in the films with the lowest residual resistivity. The O(2) deficiency model cannot explain this experimental result. Further, in Fig. 4, the dependence of the lattice constants on the reduction atmosphere argues against electron doping. It is because electron doping stretches the Cu-O bond and thereby expands $a_0$ by adding electrons to the Cu-O $dp\sigma$ anti-bonding bands [14, 15]. Actually, the $a_0$ is almost constant independent of $P_{O_2}$. In addition, it is important to emphasize that we see no indication of an expected Mott-Hubbard transition to a highly insulating state around $y \sim 0$. Namely, the films reduced in high molecular oxygen pressure lose superconductivity, but stay metallic above 150 K. Our oxygen nonstoichiometory experiments indicate that the behaviour in $T'$-(La,$RE$)$_2$CuO$_{4+y}$ cannot be explained by progressive carrier doping due to increased oxygen deficiencies by lowering $P_{O2}$ in reduction, but it can be explained in the other way around, namely, by progressive carrier localization due to increased impurity oxygen atoms by increasing $P_{O2}$ in reduction. The impurity oxygen at apical sites seems to cause strong magnetic impurity scattering by Cu$3d$ spins induced just beneath apical oxygen atoms.

## 4. Summary

We have systematically investigated the oxygen nonstoichiometry effect on $T'$-La$_{1.85}$Y$_{0.15}$CuO$_{4+y}$. The superconducting properties are strongly affected by the oxygen nonstoichiometry. The highest $T_c$ and lowest resistivity are simultaneously achieved with the optimum reduction. Either insufficient reduction or excessive reduction reduces $T_c$ and also increases resistivity. An upturn in resistivity is observed in the case of insufficient *reduction*, but not in the case of excessive *reduction*. We did not see a maximum of resistivity at "hypothetical" zero doping for a Mott insulator by varing $P_{O_2}$ reduction. Our results can be systematically understood by postulating that *T'*-(La,*RE*)$_2$CuO$_{4+y}$ with $y \sim 0$ has a metallic (superconducting) ground state, which is significantly modified by the presence of apical oxygen that is a strong pair breaker and strong scatterer.

**Acknowledgements**

The authors are indeed grateful to Prof. L. Alff of Vienna University and Mr. Y. Krockenberger of Max-Plank-Institute for Solid State Research for fruitful discussions. They also thank Dr. T. Yamada, Dr. H. Sato, Dr. H. Shibata, Dr. S. Karimoto, Dr. K. Ueda, Dr. J. Kurian, and Dr. A. Matsuda for helpful discussions, and Dr. T. Makimoto and Dr. H. Takayanagi for their support and encouragement.

**REFEENCES**

[1] Y. Tokura, H. Takagi, S. Uchida, Nature (London) 337 (1989) 345.

[2] H. Takagi, S. Uchida, Y. Tokura, Phys. Rev. Lett. 62 (1989) 1197.

[3] K. Kumagai, M. Abe, S. Tanaka, Y. Maeno, T. Fujita, K. Kadowaki, Physica B 165-166 (1990) 1297.

[4] M. Matsuura, P. Dai, H. J. Kang, J. W. Lynn, D. N. Argyriou, K. Prokes, Y. Onose, Y. Tokura, Phys. Rev. B 68 (2003) 144503.

[5] M. Brinkmann, T. Rex, H. Bach, K. Westerholt, Phys. Rev. Lett. 74 (1995) 4927.

[6] A. Tsukada, Y. Krockenberger, H. Yamamoto, M. Naito, cond-mat 0311380, *ibid* 0401120.

[7] A. Tsukada, Y. Krockenberger, M. Noda, D. Manske, L. Alff, M. Naito, submitted to Phys. Rev. Lett.

[8] A. Tsukada, Y. Krockenberger, H. Yamamoto, D. Manske, L. Alff, M. Naito, Solid State Commun. 133 (2005) 427.

[9] M. Naito, H. Sato, H. Yamamoto, Physica C 293 (1997) 36.

[10] A. Tsukada, T. Greibe, M. Naito, Phys. Rev. B 66 (2002) 184515.

[11] M. Imada, A. Fujimori, Y. Tokura, Rev. Mod. Phys. 70 (1998) 1039.

[12] T. Sekitani, N. Miura, M. Naito, Int. J. Mod. Phys. B 16 (2002) 3216.

[13] T. Sekitani, M. Naito, N. Miura, Phys. Rev. B 67 (2003) 174503.

[14] Y. Tokura, H. Takagi, S. Uchida, Nature (London) 337 (1989) 345.

[15] E. Wang, J.-M. Tarascon, L. H. Greene, G. W. Hull, W. R. McKinnon, Phys. Rev. B 41 (1990) 6582.

**Figure captions**

Fig. 1. Temperature dependence of the resistivity for $La_{1.85}Y_{0.15}CuO_{4+y}$ films with different annealing temperature ($T_{red}$) (●: $T_{red}$ = 650˚C; ■: 630˚C; ♦: 600˚C; ▲: 570˚C).

Fig. 2. Temperature dependence of the resistivity for $La_{1.85}Y_{0.15}CuO_{4+y}$ films with different annealing time ($t_{red}$) (●: $t_{red}$ = 0 min; ■: 1 min; ♦: 5 min; ▲: 10 min; ○: 20 min; □: 60 min).

Fig. 3. Temperature dependence of the resistivity for $La_{1.85}Y_{0.15}CuO_{4+y}$ films with different annealing atmosphere ($P_{O_2}$ or $P_{O_3}$) (▲: $P_{O_2}$ = $10^{-8}$ Torr; ▼: $10^{-7}$ Torr; ♦: $10^{-6}$ Torr; ●: $10^{-5}$ Torr; ■: $10^{-4}$ Torr; ○: $P_{O_3}$ = $10^{-5}$ Torr; □: $10^{-4}$ Torr). Solid and broken lines represent the films annealed in oxygen ($O_2$) and ozone ($O_2$ + 10%$O_3$) atmosphere, respectively.

Fig. 4. *a*-axis and *c*-axis lattice constants ($a_0$ and $c_0$) of *T'*-$La_{2-x}Y_xCuO_4$ films as a function of annealing atmosphere ($P_{O_2}$). Closed circles and squares represent $a_0$ and $c_0$, respectively.

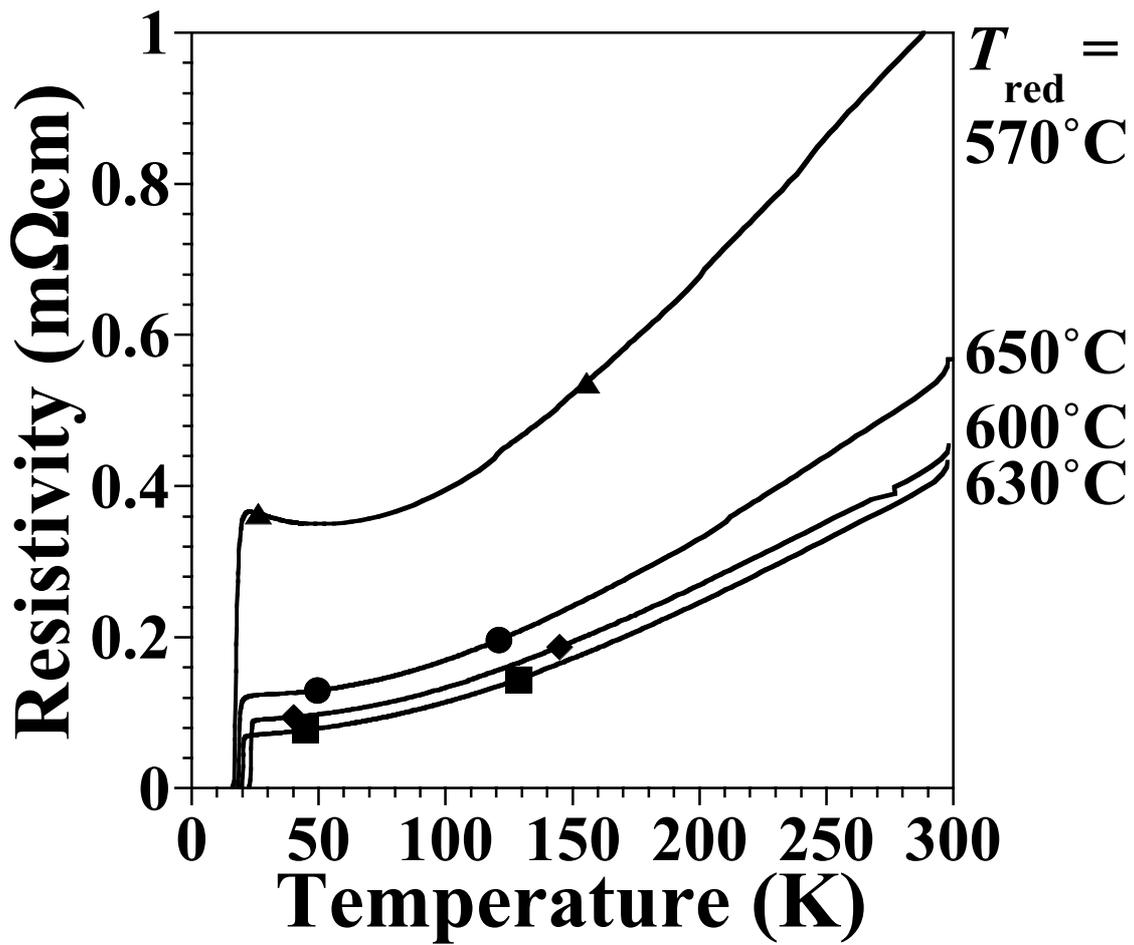

**Figure 1, Tsukada *et al.***

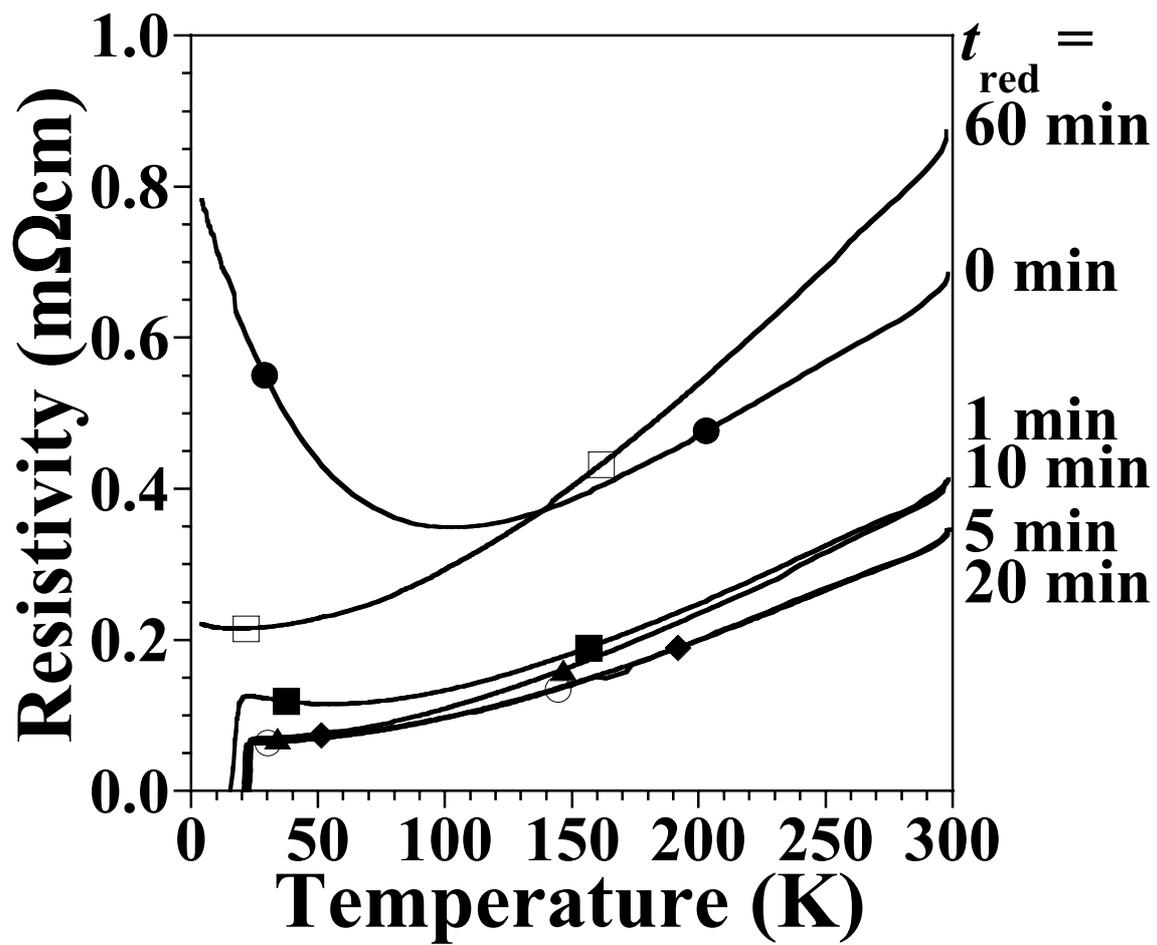

Figure 2, Tsukada *et al.*

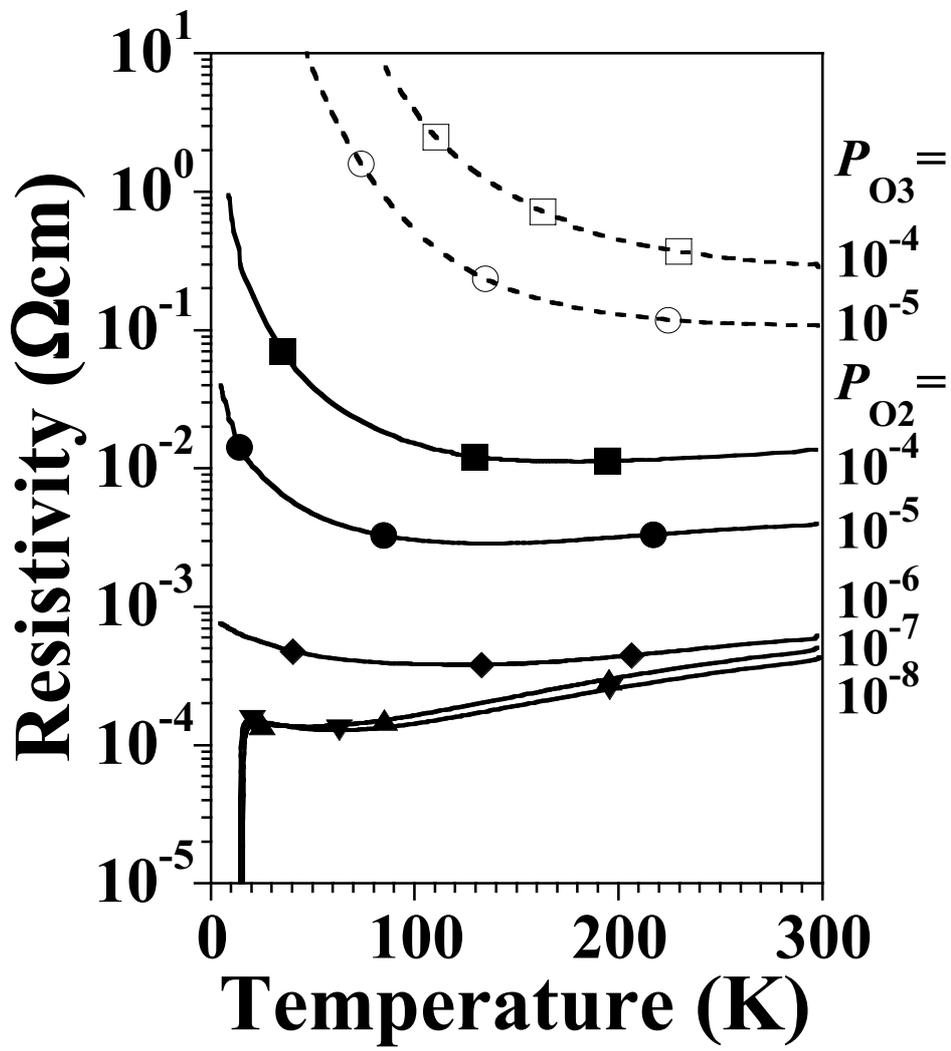

**Figure 3, Tsukada *et al.***

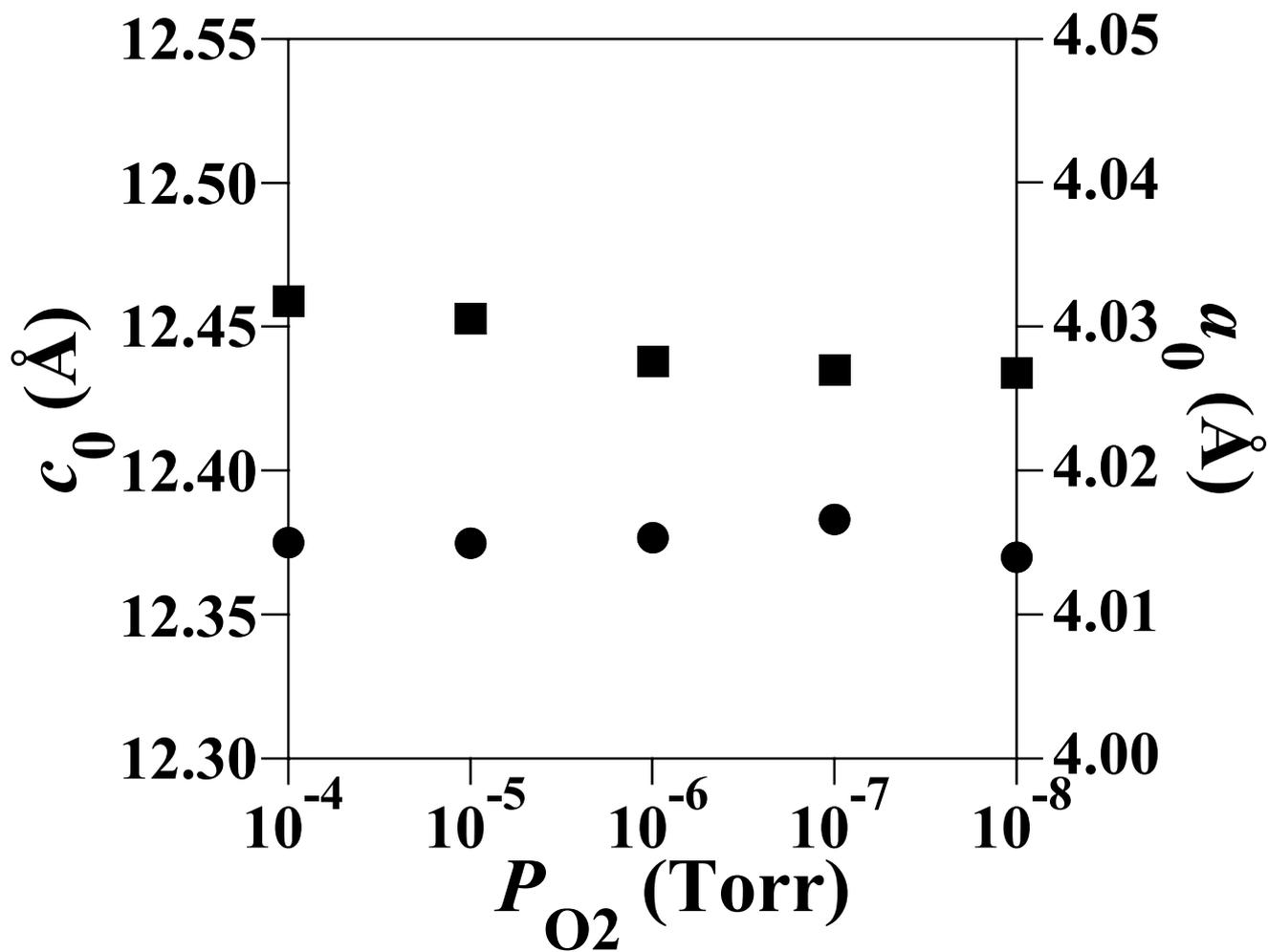

**Figure 4, Tsukada *et al.***